\documentclass[%
reprint,
amsmath,amssymb,
aps,
prb,
]{revtex4-1}
\usepackage{graphicx}
\usepackage{dcolumn}
\usepackage{bm}
\usepackage{graphicx}
\usepackage{subcaption}
\usepackage{tabularx}
\usepackage{amsmath}
\usepackage{physics}
\usepackage[hyperindex]{hyperref}
\begin{document}


\title{Ferromagnetic resonance in thin films -- cross-validation analysis of
numerical solutions of Smit-Beljers equation. Application to GaMnAs
}

\author{P.~Tomczak} 
\email{Corresponding author: ptomczak@amu.edu.pl}
	\affiliation{Quantum Physics Division\\ Faculty of Physics, Adam Mickiewicz University 
ul. Umultowska 85, 61-614 Pozna\'n, Poland}
\author{H.~Puszkarski}
\affiliation{Surface Physics Division \\ Faculty of Physics, Adam Mickiewicz University 
ul. Umultowska 85, 61-614 Pozna\'n, Poland} 

\date{\today}

\begin{abstract}
The new method of numerical analysis of experimental ferromagnetic resonance
(FMR) spectra in thin films is developed and applied to
(Ga,Mn)As thin films. 
Specifically, it 
starts with the finding of numerical solutions of Smit-Beljers
(SB) equation and continues with their subsequent statistical analysis
within the cross-validation (CV) approach
taken from machine learning techniques.
As a result of this treatment we are able to reinterpret the available
FMR experimental results in diluted ferromagnetic semiconductor (Ga,Mn)As thin films
with resulting determination of 
magnetocrystalline anisotropy 
constants.
The outcome of CV analysis points that it is necessary
to take into account terms describing the bulk cubic anisotropy up to the fourth order
to reproduce FMR experimental results for (Ga,Mn)As correctly.
This finding contradicts the wide-spread conviction in the literature
that only first order cubic anisotropy term is important in this material.
We also provide numerical values of these higher order cubic anisotropy constants 
for (Ga,Mn)As thin films resulting from SB-CV approach.
\end{abstract}

\pacs{Valid PACS appear here}
\maketitle
\section{\label{Intro}Introduction: The Experimental Data}
Gallium manganese arsenide, (Ga,Mn)As, is probably one of the most thoroughly 
studied diluted ferromagnetic semiconductors.
Simultaneous presence of magnetism and conductivity in this material
makes it possible to
control of both the charge and the spin degrees of freedom of the charge carriers.
This creates potential spintronic applications.
Another reason for the intense research on (Ga,Mn)As
are its remarkable magnetic properties between which magnetic anisotropy plays an important role.
It determines, among others, the orientation of magnetization in the absence 
of an applied magnetic field\cite{Welp2003}. 
Although its understanding 
is important for  prospective applications such as e.g., memory devices,
its origins  are far from being fully explained.
Magnetocrysralline  anisotropy in (Ga,Mn)As,
usually described by the single-domain model\cite{Liu2003, Nie2010},
is being investigated by various 
experimental techniques, such as ferromagnetic resonance (FMR) 
and spin-wave resonance (SWR)\cite{Liu2006}.
Most of these methods have been used to obtain information on anisotropy bulk properties
of this material.
Recently we have proposed\cite{Pusz2017} that one can use the SWR to 
get information on such magnetic properties as surface anisotropy 
and surface pinning energy of (Ga,Mn)As thin films and their dependence 
on the orientation of magnetization in the material.  

\begin{figure}[!ht]
\centering
   \includegraphics[width=0.20\textwidth, angle=0]{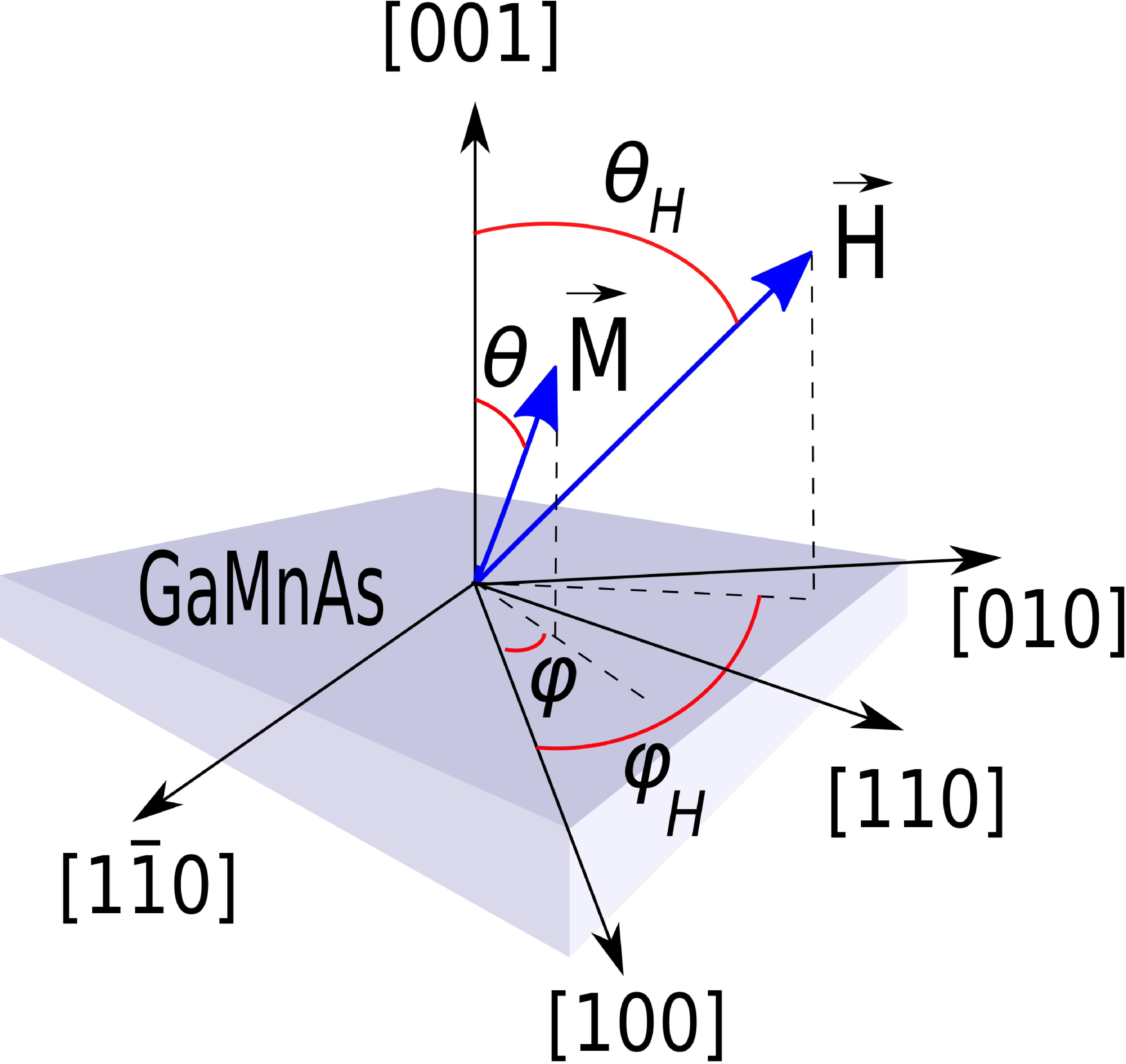}
   \caption{
   The coordinate system in which an orientation of the applied
   magnetic field~${\mathbf H}$ is described
   with respect to the sample in the FMR experiment.
   The field direction is characterized by angles
$\vartheta_H$ and $\varphi_H$ measured relative to the sample [001] and [100] axes.
The equilibrium direction of the sample     
magnetization~${\mathbf M}$ is represented by angles $\vartheta$ and $\varphi$.}
   \label{sample_geometry}
\end{figure}

The ferromagnetic resonance spectroscopy has long been a good tool for
examining magnetocrysralline anisotropy, see e.g.,
recent review on FMR in (Ga,Mn)As thin films\cite{Liu2006}.
In FMR experiment, 
since the equilibrium position of 
the total magnetic
moment $\mathbf M$ of the sample 
does not coincide with the direction 
of magnetic field $\mathbf H$ due to the presence of magnetocrysralline
anisotropy,  
$\mathbf M$ precesses around its equilibrium position
with a specific (microwave) frequency $\omega$.
By changing the magnetic field $\mathbf H$ one hits a resonance field $\mathbf H_r$:
the precession frequency of $\mathbf M$ is equal to the frequency of the spectrometer.
The value of the resonance
field $\mathbf H_r$ strongly depends on its orientation with respect 
to the examined sample, which is
determined by angles $\theta_H$ and $\phi_H$, see Fig.~\ref{sample_geometry},
due
to magnetocrysralline anisotropy.

This article presents the results of the analysis of bulk magnetocrystalline anisotropy in (Ga,Mn)As
based on examination of the uniform mode in SWR resonance in (Ga,Mn)As 
thin film\cite{Liu2007}. The motivation to carry out this 
analysis was two-fold: First --  on the basis of examination of surface mode
in the same SWR experiment\cite{Liu2007}
we have schown\cite{Pusz2017} that magnetocrystalline $sur\!f\!ace$ anisotropy  
in (Ga,Mn)As thin films contains cubic terms up to third order, which is not commonly found among ferromagnets.
We wonder if this is also true for $bulk$ magnetocrystalline 
anisotropy.
Second -- it was originally shown\cite{Liu2007} that in order to reproduce the experimental
dependence of the resonance field on the orientation of the magnetic field
with respect to the sample, only the first order term of cubic anisotropy should be taken into account,
which, to some extent, is contradictory
to the analysis carried out for the surface\cite{Pusz2017}.
In the meantime, numerical tools have been developed that allow a thorough analysis of this problem.
That is why we have considered the old problem again.

\begin{figure}
\centering
   \includegraphics[width=0.35\textwidth, angle=0]{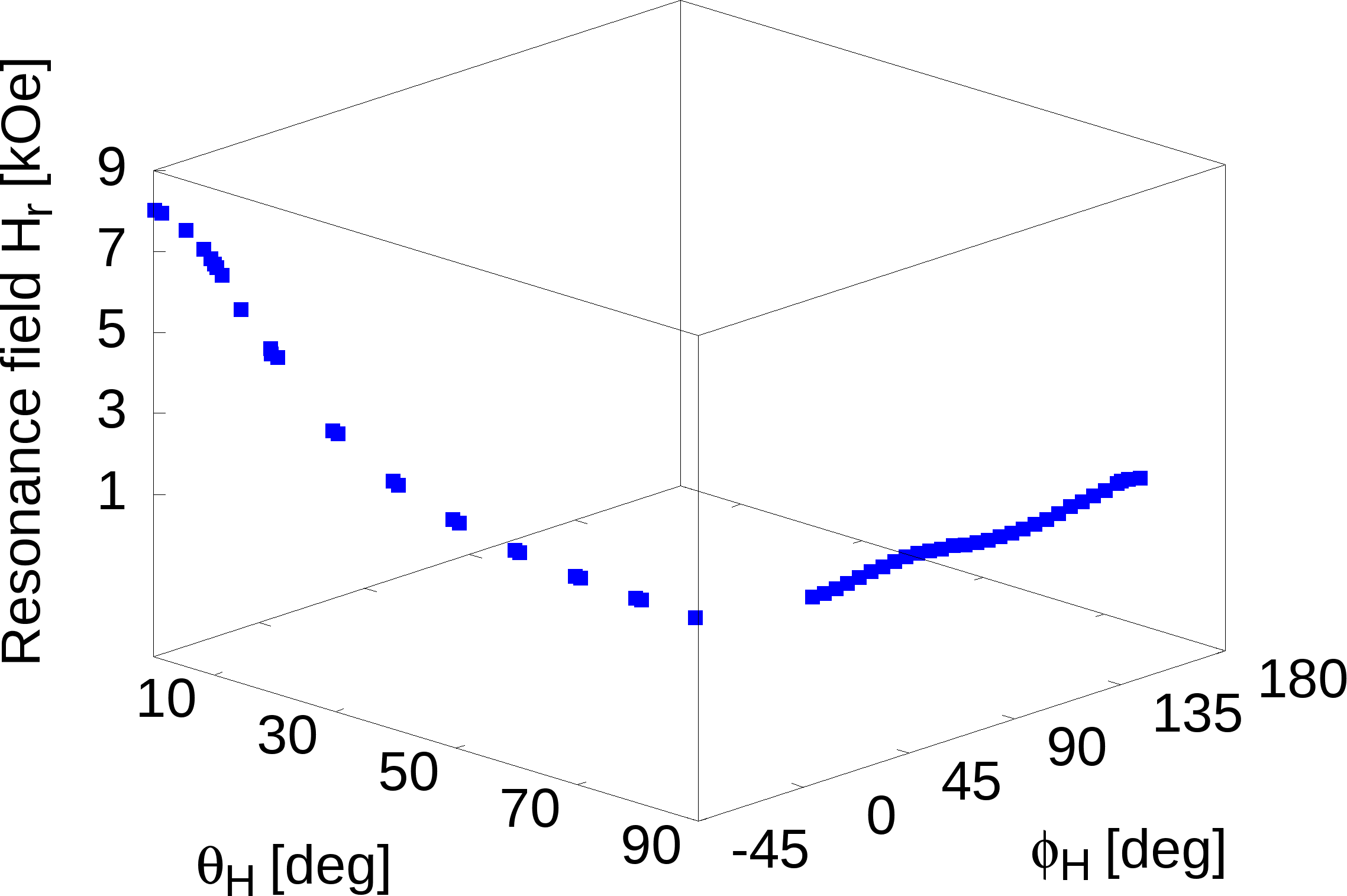}
   \caption{Resonance field\cite{Liu2007} $H_r$ of the uniform SWR mode 
            as a function of the magnetic field orientation for the
            out-of-plane configuration (plane $H_r$-$\vartheta_H$) and for the
            in-plane configuration (plane $H_r$-$\varphi_H$).
}
\label{experimental_data}
\end{figure}

At the begining let us recall the 
angular dependencies of resonance field
for the uniform mode 
in ferromagnetic
(Ga,Mn)As thin film\cite{Liu2007}.
They are shown in Ref.~[\onlinecite{Liu2007}]
in Fig.\,5 for the out-of-plane geometry and in Fig.\,6 for the
in-plane geometry, respectively.
We show them again in Fig.~\ref{experimental_data}, 
to clearly emphasize the difference between resonance field resulting from the uniaxial anisotropy (plane $H_r$-$\vartheta_H$)  
and that resulting from the cubic anisotropy (plane $H_r$-$\varphi_H$). 
We focus on the interpretation of this experiment because
the authors very carefully identified 
resonance from uniform SWR modes
and distinguished it from that for surface modes.
\section{\label{Free}Phenomenological free energy}

\begin{figure}
\centering
  \begin{subfigure}[b]{.2\linewidth}
    \centering
    \includegraphics[width=.75\textwidth]{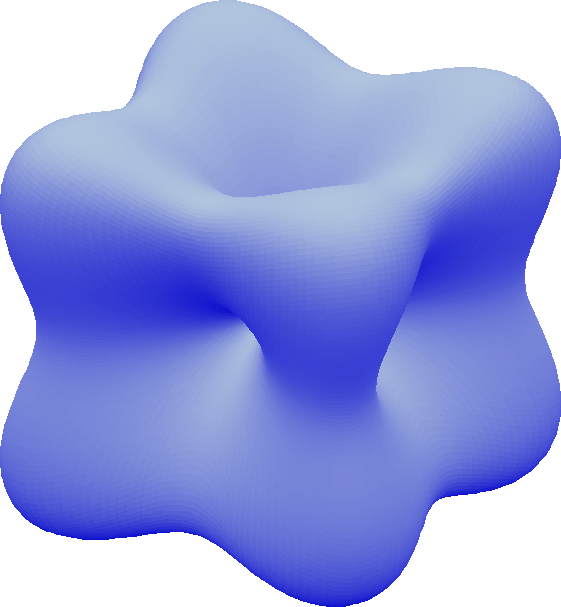}
    \caption{}\label{fig:1a}
  \end{subfigure}%
  \qquad \qquad
  \begin{subfigure}[b]{.2\linewidth}
    \centering
    \includegraphics[width=.75\textwidth]{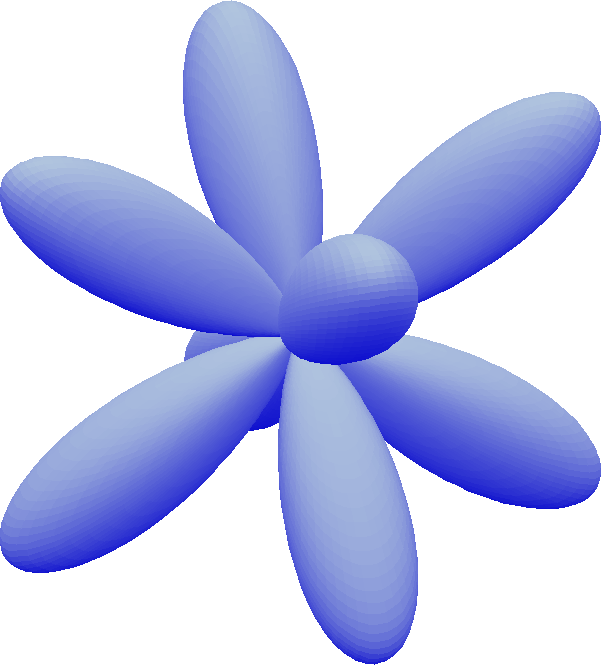}
    \caption{}\label{fig:1b}
  \end{subfigure}%
  \qquad \qquad
  \begin{subfigure}[b]{.2\linewidth}
    \centering
    \includegraphics[width=.75\textwidth]{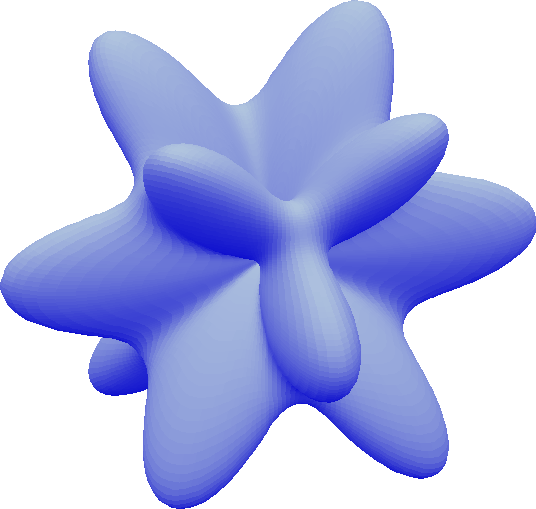}
    \caption{}\label{fig:1c}
  \end{subfigure}  
  \begin{subfigure}[b]{.2\linewidth}
    \centering
    \includegraphics[width=.75\textwidth]{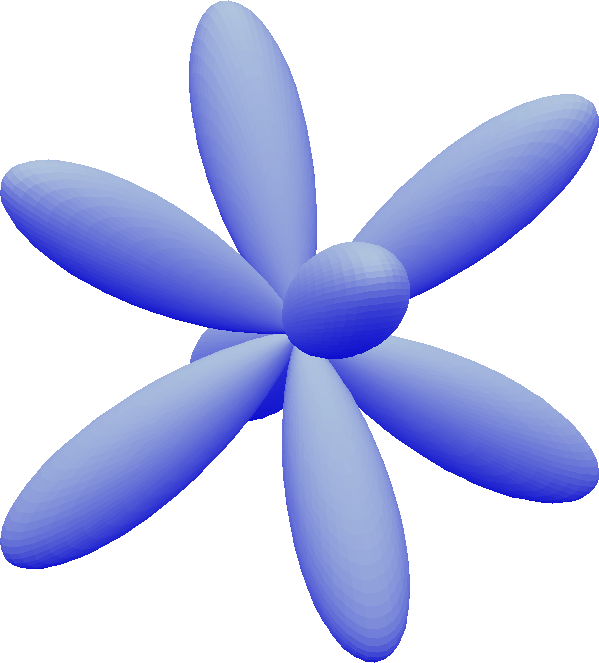}
    \caption{}\label{fig:1d}
  \end{subfigure}%
  \qquad \qquad
  \begin{subfigure}[b]{.2\linewidth}
    \centering
    \includegraphics[width=.65\textwidth]{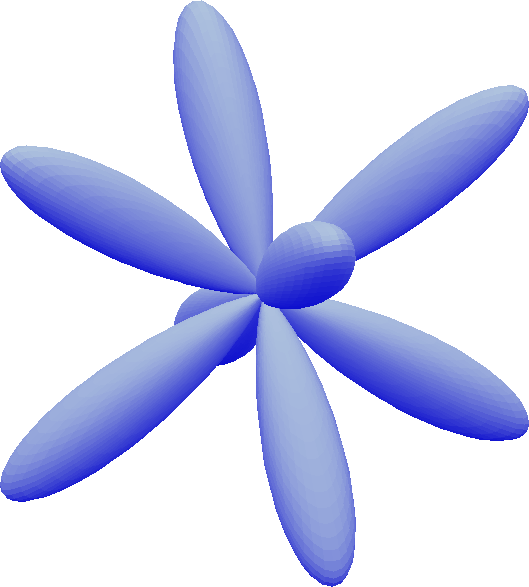}
    \caption{}\label{fig:1e}
  \end{subfigure}%
  \qquad \qquad
  \begin{subfigure}[b]{.2\linewidth}
    \centering
    \includegraphics[width=.6\textwidth]{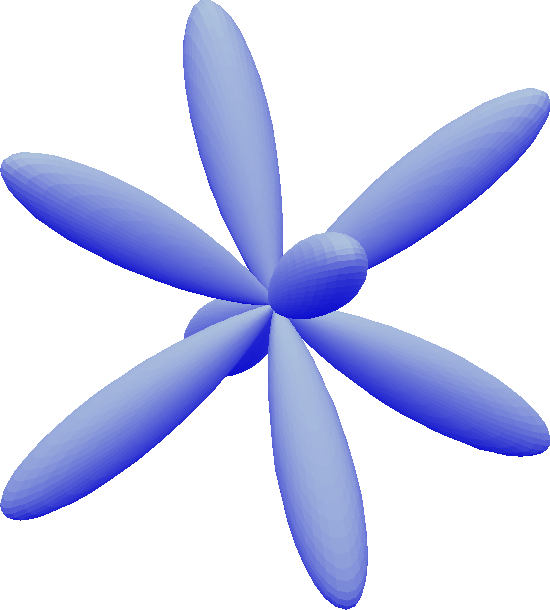}
    \caption{}\label{fig:1f}
  \end{subfigure}%
  \caption{
  The basis functions up to 6$^{\mbox{th}}$ order (a-f) used in the expansion (\ref{cubic_eq})
  of the cubic magnetocrystalline field. Although each next function of a higher order is, 
  for given $ \vartheta, \varphi$, smaller than the preceding one,
  they are shown here as being comparable in size.}
\label{cub_terms}
\end{figure}

The starting point for the interpretation of experimental data
of ferromagnetic resonance in (Ga,Mn)As 
is the phenomenological formula for the free energy of the 
investigated sample.
We assume that
there exists a single
homogeneous magnetic domain within the sample and that the
free energy of unit volume of the sample consists of Zeeman term $F_Z$, demagnetization term $F_D$, 
and magnetocrystalline anisotropy terms (cubic $F_C$ and uniaxiall $F_U$):
\begin{equation}
\label{free_energy}
F = F_Z + F_D + F_C + F_U.
\end{equation}
Expressing free energy in terms of {\em fictitious} fields one obtains
\begin{equation}
f(\vartheta_H, \varphi_H, \vartheta, \varphi)=\frac{F}{M} = H_Z + H_D + H_C + H_U,
\label{field_eq}
\end{equation}
$M$ stands here for the value of homogeneous magnetization.
Dependence of fictitious anisotropy fields
in Eq.~(\ref{field_eq})
on the direction in space is expressed by
unit vectors
along the applied
magnetic field $\mathbf H$ and along the magnetization of the sample $\mathbf M$.
Their spatial orientation is determined by angles $ \vartheta_H$, $ \varphi_H$ 
and $ \vartheta$, $ \varphi$ with respect to the [001] and [100] axes,
see Fig. \ref{sample_geometry}.
The unit vectors
are given by 
\begin{subequations}
\begin{align}
\frac{\mathbf H}{H} = [n^H_x,  n^H_y,  n^H_z] = 
[\cos{\varphi_H}\sin{\vartheta_H}, \sin{\varphi_H}\sin{\vartheta_H}, \cos{\vartheta_H}], \\
\frac{\mathbf M}{M} = [n_x,  n_y,  n_z] = 
[\cos{\varphi}\sin{\vartheta}, \sin{\varphi}\sin{\vartheta}, \cos{\vartheta}].
\end{align}
\end{subequations}
Zeeman field is given by
 \begin{equation}
H_Z(\vartheta_H, \varphi_H, \vartheta, \varphi)  = - H ( n_x n^H_x + n_y n^H_y + n_z n^H_z ).
\label{Zeeman_eq}
\end{equation}
Demagnetization field of a thin film
may be approximated by the expression
describing demagnetization field of an infinite plane
 \begin{equation}
H_D(\vartheta) =  2 \pi M n_z^2.
\label{demag_eq}
\end{equation}
The field $H_C(\vartheta, \varphi)$ should be invariant 
under the 
cubic symmetry
transformations.
It follows that it is possible
to expand it into basis functions with the same symmetry.
Typically this expansion is limited to some low-order terms 
of systematically decreasing basis functions.
The way of constructing such basis functions is presented, e.g., in Ref.~[\onlinecite{Zemen2009}]:
they are chosen
from all terms of the expansion of the identity $(n^2_x + n^2_y + n^2_z)^n=1$ ($n=2,3...$)
which are invariant under
permutation of the indices $x$, $y$, and $z$.
Expansion up to $n=7$ is used in what follows:
\begin{equation}
\begin{split}
 H_C(\vartheta, \varphi)= H_{c1}(n^2_x n^2_y + n^2_y n^2_z + n^2_z n^2_x) + \\ 
 H_{c2}(n^2_x n^2_y n^2_z) + \\
 H_{c3}(n^4_x n^4_y + n^4_y n^4_z + n^4_z n^4_x) + \\
 H_{c4}(n^4_x n^4_y n^2_z + n^4_x n^2_y n^4_z + n^2_x n^4_y n^4_z)+\\
 H_{c5}(n^4_x n^4_y n^4_z) + \\
 H_{c6}(n^6_x n^6_y n^2_z + n^6_x n^2_y n^6_z + n^2_x n^6_y n^6_z).
\label{cubic_eq}
\end{split}
\end{equation}
All terms included in the expansion of cubic field $H_C(\vartheta, \varphi)$ are shown in Fig. \ref{cub_terms} (a)-(f).
Let us emphasize that every next term is smaller than the previous one - the expansion (\ref{cubic_eq}) is convergent.
Let us note, however, that the set of six basis functions used in expansion (\ref{cubic_eq}) is neither orthogonal nor complete.
Consequently, the expansion may not be unique\cite{Callen1960}.
Nevertheless, 
it is widely used,
at least in the cases of analyzing systems with lower order magnetocrystalline anisotropies.
One can also expand the cubic magnetocrystalline field $ H_C(\vartheta, \varphi)$ into other basis functions, such as, e.g., spherical harmonics,
remembering however,  to use only those with the appropriate symmetry\cite{Pusz1979}.

\begin{figure}
\centering
\begin{subfigure}[b]{.09\linewidth}
    \centering
    \includegraphics[width=.75\textwidth]{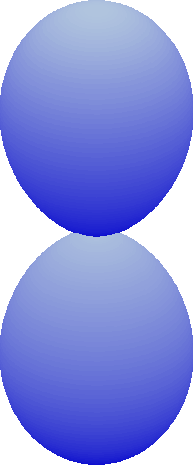}
    \caption{}\label{fig:2a}
  \end{subfigure}%
  \qquad \qquad
  \begin{subfigure}[b]{.07\linewidth}
    \centering
    \includegraphics[width=.75\textwidth]{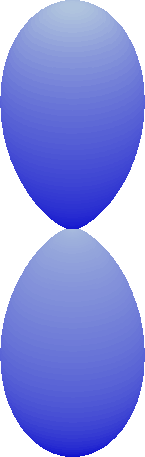}
    \caption{}\label{fig:2b}
  \end{subfigure}%
  \qquad \qquad
  \begin{subfigure}[b]{.055\linewidth}
    \centering
    \includegraphics[width=.75\textwidth]{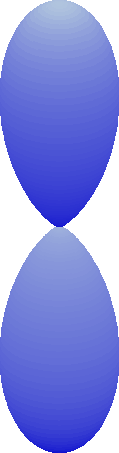}
    \caption{}\label{fig:2c}
  \end{subfigure}%
  \qquad \qquad
  \begin{subfigure}[b]{.18\linewidth}
    \centering
    \includegraphics[width=.82\textwidth]{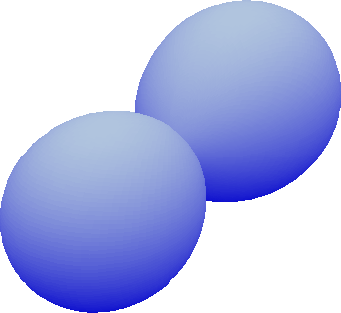}
    \caption{}\label{fig:2d}
  \end{subfigure}%
  \caption{
  Terms representing contributions to uniaxial magnetocrystalline anisotropy
  of (Ga,Mn)As thin film
  in spherical coordinate system.
  Uniaxial anisotropy along $z$ axis 1$^{\mbox{st}}$ to 3$^{\mbox{rd}}$ order:
  $n^2_z$ - (a), $n^4_z$ - (b), $n^6_z$ - (c).
  Uniaxial anisotropy along $[110]$ axis $(n_y - n_x)^2$ - (d).}
  \label{uni_terms}
\end{figure}

It is recognized\cite{Liu2006, Zemen2009} that uniaxiall anisotropy field $H_U(\vartheta, \varphi)$ in (Ga,Mn)As consist 
of two terms $H_{1[001]}n^2_z$  and $H_{[110]}(n_y - n_x)^2$ representing 
uniaxial anisotropy along $z$ axis and uniaxial anisotropy along [110] axis, respectively.
We add, however, two additional terms of 4$^{\mbox{th}}$ and  6$^{\mbox{th}}$ order:
\begin{equation}
\begin{split}
H_U(\vartheta, \varphi)= -\frac{1}{2}H_{1[001]}n^2_z - \frac{1}{2}H_{2[001]}n^4_z - \frac{1}{2}H_{3[001]}n^6_z \\
- \frac{1}{2}H_{[110]}(n_y - n_x)^2.
\label{UNI}
\end{split}
\end{equation}
Note that two terms with $\cos^2{\vartheta_H}$ are present in expansion \ref{field_eq}: $2\pi M$ 
and $-\frac{1}{2}H_{1[001]}$.
Usually they are grouped together and referred to as effective anisotropy field: 
$H_{[001]}^{ef\!f}=2\pi M - \frac{1}{2}H_{1[001]})$.
Four terms entering the expansion of uniaxial field $H_U(\vartheta, \varphi)$ are shown in Fig. \ref{uni_terms}.

The right side of Eq.~(\ref{field_eq}) depends on 15 variables:
$H_r,\vartheta_H,\varphi_H,\vartheta,\varphi,H_{c1},...,H_{c6},H_{[001]}^{ef\!f},H_{2[001]},H_{3[001]},\\H_{[110]}$.
The first three will be considered as independent variables that are measured in the experiment and are
shown in Fig. \ref{experimental_data}.
The pairs of angles $(\vartheta_H, \varphi_H)$ 
and $(\vartheta, \varphi)$ are not independent.
The angles determining the equilibrium orientation of $\mathbf M$, $\vartheta$ and $\varphi$, should minimize
free energy.
For the set of fixed parameters $H_r,\vartheta_H,\varphi_H,H_{c1},...,H_{c6},H_{[001]}^{ef\!f},H_{2[001]},H_{3[001]},H_{[110]}$,
one finds them from the equilibrium condition
\begin{equation}
\frac{\partial f}{\partial \vartheta}=0, \,\,\, \frac{\partial f}{\partial \varphi}=0.
\label{equilibrium_eq}
\end{equation}
Thus the right-hand side of Eq.~(\ref{field_eq}) really depends on 10 fictitious
anisotropy fields
$H_{c1},...,H_{c6},H_{[001]}^{ef\!f},...,H_{3[001]},H_{[110]}$ 
which we collectively denote by the vector 
\begin{equation}
{\mathbf h} \equiv  (H_{c1},...,H_{c6},H_{[001]}^{ef\!f},H_{2[001]},H_{3[001]},H_{[110]}).
\label{v_fields}
\end{equation}
The questions should then be asked:
how to find anisotropy fields and how many 
of them are necessary to reproduce the experimental dependence $H_r(\vartheta_H, \varphi_H)$ well?
These questions are answerd in the next section. ￼

\section{\label{Resonance} 
Which anisotropy fields are important for (Ga,Mn)as?}

The resonance condition (in the case of uniform magnetization) is given by \cite{Smit55, Wigen88}
 \begin{equation}
\bigg (\frac{\omega}{\gamma} \bigg)^2 = \frac{1}{\mbox{sin}^2\vartheta}(f_{\vartheta \vartheta} f_{\phi \phi} - f^2_{\vartheta \phi}),
\label{SB_eq}
\end{equation}
where $f_{\vartheta \phi}=\frac{\partial f}{\partial \vartheta}\frac{\partial f}{\partial \varphi}$,
$\gamma=\frac{g\mu_B}{\hbar}$ with $g$ being the spectroscopic splitting factor,
$\mu_B$ the Bohr magneton and $\hbar$ - the Planck constant.
The 9.46 GHz spectrometer was used in the considered experiment\cite{Liu2007}, thus Eq.~(\ref{SB_eq}) reads
\begin{equation}
45.6834 = \frac{g^2}{\mbox{sin}^2\vartheta}(f_{\vartheta \vartheta} f_{\phi \phi} - f^2_{\vartheta \phi}),\,\,\,[\mbox{kOe}^2].
\label{SB_num_eq}
\end{equation}

At resonance Eq.~(\ref{SB_num_eq}) should be met for any given direction of $\mathbf H_r(\vartheta_H, \varphi_H)$,
i.e., in the case under consideration, for all points shown in Fig. \ref{experimental_data}.
Let us treat $g$ and components of the vector ${\mathbf h}$ as not known parameters and denote
the right-hand side of the Eq.~(\ref{SB_num_eq}) calculated at $i$-th experimental point by $R_i$,
\begin{equation}
R_i(g,{\mathbf h} ) = 
\frac{g^2}{\mbox{sin}^2\vartheta}(f^i_{\vartheta \vartheta}
f^i_{\phi \phi} - (f^i_{\vartheta \phi})^2).
\label{Left_SB_num_eq}
\end{equation}
Note that free energy derivatives (e.g., $f^i_{\vartheta \vartheta}$) are calculated at $i$-th the experimental point
--- for given values $H_r,\vartheta_H,\varphi_H$.

One should find such values of unknown coefficients $g,{\mathbf h}$  that the Eq.~(\ref{SB_num_eq}) 
is met as accurately as possible
for each experimental point. 
So the following error function, being the positive
 square root of the sum of squares of residuals,
\begin{equation}
E^N_{RMS}(g,{\mathbf h} ) = \sqrt{\frac{1}{N}\sum_i \big(R_i(g,{\mathbf h}) - 45.6834\big)^2},
\label{Opt_num_eq}
\end{equation}
should be minimized
in 11-dimensional parameter space $(g,{\mathbf h})$.
The sum in Eq.~(\ref{Opt_num_eq}) runs over all $N$ experimental points shown in Fig. \ref{experimental_data}.
This least squares approach to finding 
the unknown parameters represents a specific case of {\em maximum likelihood} approach\cite{Bishop_2006,Mehta2018} .
Actual values of the spectroscopic splitting factor $g$ and magnetocrysralline anisotropy fields
are those for which Eq.~(\ref{Opt_num_eq}) has a minimum
close to zero.

We have 
included 10 magnetocrysralline anisotropy fields into
the formula for free energy.
Now we will check which ones are really essential
to describe the experimental results well
using a simple {\em cross-validation} scheme\cite{Bishop_2006}.
For this purpose anisotropy models are defined
in Tables \ref{Cubic_present} and \ref{Uni_present}.
For example, in the model C3 (third row of Table \ref{Cubic_present})
the cubic anisotropy is expanded up to 3$^{\mbox{rd}}$ order, 
the uniaxial anisotropy along $z$ axis up to 1$^{\mbox{st}}$ order,
the uniaxial anisotropy along [110] axis up to 1$^{\mbox{st}}$ order
and similarly for other models.
\begin{table}
\caption{
Models of cubic magnetocrystalline anisotropy 
in (Ga,Mn)As
for which cross-validation was carried out.
In the second column the terms are given 
included in the expansion of the free energy (\ref{cubic_eq}) 
for models C1 - C6.
Uniaxial anisotropy for C1 - C6 models does not change ---
only fields $H_{[001]}^{ef\!f}$ and $H_{[110]}$ are present.}
\begin{ruledtabular}
\begin{tabular}{ccc}
\\
Model&Cubic anisotropy&Uniaxial anisotropy\\
\\
\hline
\\
C1 & $H_{c1}$       & $H_{[001]}^{ef\!f},H_{[110]}$\\
C2 & $H_{c1},H_{c2}$& $H_{[001]}^{ef\!f},H_{[110]}$\\
C3 & $H_{c1}-H_{c3}$& $H_{[001]}^{ef\!f},H_{[110]}$\\
C4 & $H_{c1}-H_{c4}$& $H_{[001]}^{ef\!f},H_{[110]}$\\
C5 & $H_{c1}-H_{c5}$& $H_{[001]}^{ef\!f},H_{[110]}$\\
C6 & $H_{c1}-H_{c6}$& $H_{[001]}^{ef\!f},H_{[110]}$\\

\\
\end{tabular}
\end{ruledtabular}
\label{Cubic_present}
\end{table}

\begin{table}
\caption{
Models of uniaxial magnetocrystalline anisotropy 
in (Ga,Mn)As
for which cross-validation was carried out.
In the second column the terms are given 
included in the expansion of the free energy (\ref{uni_terms})
for models U1 - U3.
Cubic anisotropy for U1 - U3 models does not change ---
only fields $H_{c1}-H_{c4}$ are present.}
\begin{ruledtabular}
\begin{tabular}{ccl}
\\
Model&Uniaxial anisotropy&Cubic anisotropy\\
\\
\hline
\\
U1 & $H_{[001]}^{ef\!f},H_{[110]}$                     & $H_{c1}-H_{c4}$ \\
U2 & $H_{[001]}^{ef\!f},H_{2[001]},H_{[110]}$          & $H_{c1}-H_{c4}$ \\
U3 & $H_{[001]}^{ef\!f},H_{2[001]},H_{3[001]}H_{[110]}$& $H_{c1}-H_{c4}$ \\
\\
\end{tabular}
\end{ruledtabular}
\label{Uni_present}
\end{table}

The cross-validation, within {\em leave-one-out technique}, runs as follows:
We divide the $N$ ($=55$) element set of experimental data into two 
subsets: the training one and the test one. The first one contains $N-1$
elements, the second one --- 1 element. 
One can do it in $N$ possible ways.
Subsequently the $N$ subsets obtained in this way 
are used to train, i.e., 
to determine the values of the unknown parameters $(g,{\mathbf h})$ 
by minimizing the error function $E^{N-1}_{RMS}(g,{\mathbf h} )$, defined 
in Eq.~(\ref{Opt_num_eq}) for each model under consideration.
Simultaneously
the error function $E_{RMS}^{1}(g,{\mathbf h} )$ is calculated for one left test point
for each model.
Note that its value informs us 
how well we are doing in predicting the values of anisotropy fields for a particular model.
After $N$ minimizations one examines how averages $\left <E_{RMS}^{N-1} \right >$ and $\left <E_{RMS}^{1}\right >$
depend on the model, i.e., on the order od expansion (\ref{cubic_eq}) or (\ref{UNI}).
We use the following criterion to assess the quality of the model:
the model describes magnetocrystalline anisotropy well if taking into account higher order terms
in the expansions (\ref{cubic_eq}) and (\ref{UNI}) does not improve its predictive ability given 
by the average $\left < E^{1}_{RMS}\right >$.

The procedure described above allowed to find the average values of anisotropy fields, $g$-factor
and error functions $\left<E_{RMS}^{N-1}\right>$ and $\left<E_{RMS}^{1}\right>$
for each model after N minimizations
for the sample in the experiment under consideration. 
They are collected in Tables \ref{Aniso_Fields} and \ref{Uni_Fields}.
\begin{figure}
\centering
   \includegraphics[width=0.35\textwidth, angle=0]{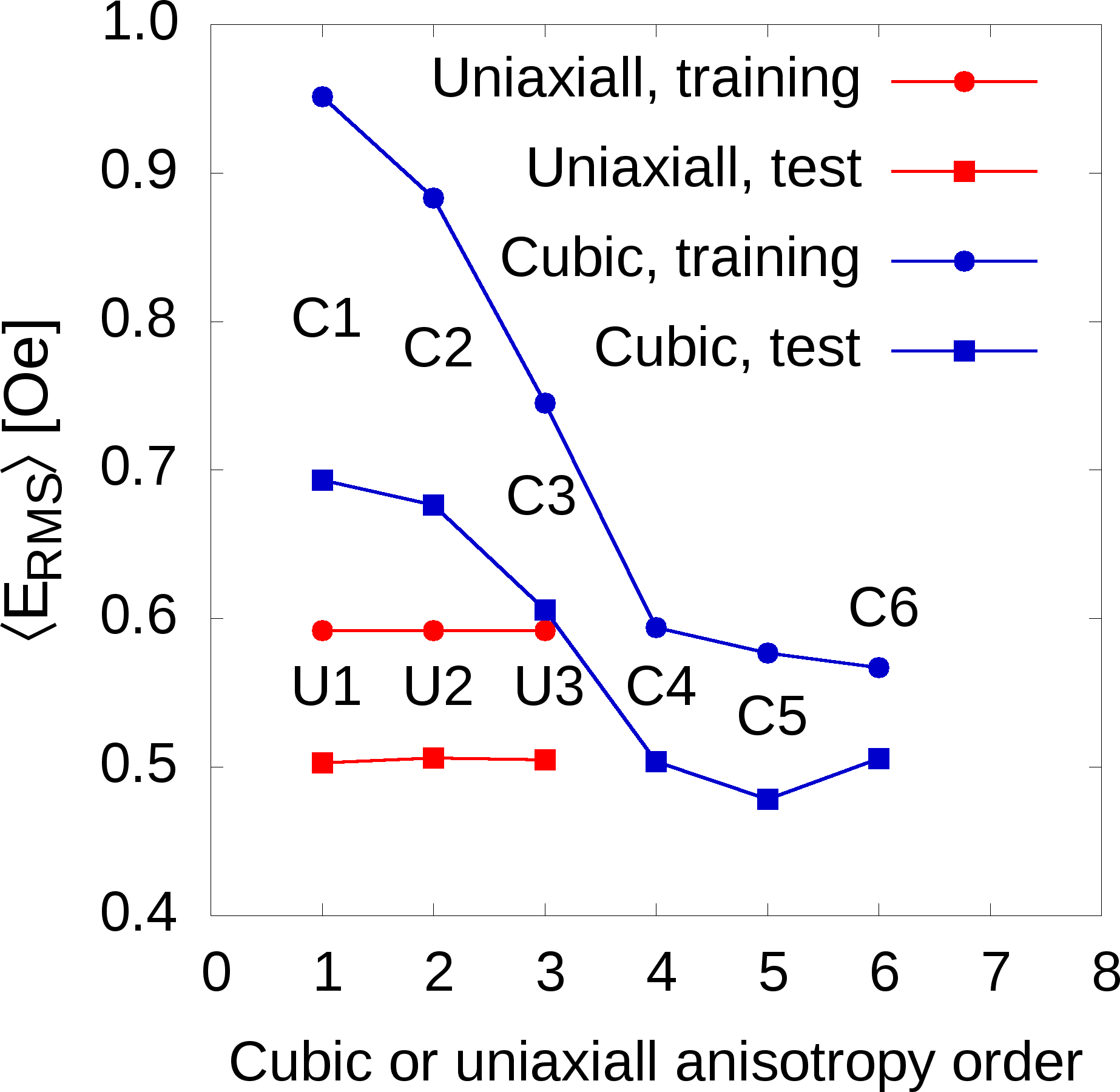}
   \caption{
   The values of error functions $\left <E_{RMS}^{N-1} \right >$ and $\left <E_{RMS}^{1}\right >$
   defined in Eq.~(\ref{Left_SB_num_eq}) for all models defined in Tables \ref{Cubic_present} and \ref{Uni_present}.
   The values error functions 
   were calculated for the corresponding field values taken from the Tables \ref{Aniso_Fields} and \ref{Uni_Fields}. }
   \label{Cr_Va_fig}
\end{figure}
\begin{table*}
\caption{Anisotropy fields [Oe] in bulk (Ga,Mn)As related to cubic and uniaxial symmetry 
and values of $g$-factor calculated for models C1-C6 according to the procedure described in the text.
In the last two columns error functions $\left<E_{RMS}^{N-1}\right>$ and $\left<E_{RMS}^{1}\right>$ 
are shown.}
\begin{tabularx}{\textwidth}{XXXXXXXXXXXX}
\hline\hline\\
Model & $H_{c1}$&$H_{c2}$& $H_{c3}$&$H_{c4}$&$H_{c5}$&$H_{c6}$& $H_{[001]}^{ef\!f}$ 
&$H_{[110]}$&$g$&$\left<E_{RMS}^{N-1}\right>$&$\left<E_{RMS}^{1}\right>$ \\
     \\
     \hline
     \\
     C1  & 91.81 &        &       &       &      &      & 4765 & 65.53 & 1.978 & 0.95 & 0.69\\
     C2  & 92.21 & -87.94 &       &       &      &      & 4774 & 70.72 & 1.979 & 0.88 & 0.68\\
     C3  & 77.06 & -4.241 & 57.00 &       &      &      & 4776 & 61.70 & 1.982 & 0.75 & 0.61\\
     C4  & 78.07 & -534.1 & 43.93 & 1405  &      &      & 4811 & 66.29 & 1.985 & 0.59 & 0.50\\
     C5  & 79.57 & -583.4 & 41.68 & 1790  & 2080 &      & 4814 & 64.50 & 1.984 & 0.58 & 0.48\\
     C6  & 78.78 & -419.0 & 41.82 & 436.1 & 2841 & 2700 & 4809 & 63.81 & 1.984 & 0.57 & 0.51\\
Liu~et.~al\cite{Liu2007}& 197\footnote{Note that 
definition  of cubic anisotropy in  Ref. [\onlinecite{Liu2007}] is slightly different.
It takes into account tetragonal distortion in (GaMn)As thin films. The 
value of cubic anisotropy field calculated from this definition
should be approximately equal to $2H_{c1}$. }
                  &       &       &       &      &      & 4588 & 77    & 1.98  &       &\\
\\
  \hline\hline
\end{tabularx}
\label{Aniso_Fields}
\end{table*}
\begin{table*}
\caption{Anisotropy fields [Oe] in bulk (Ga,Mn)As related to cubic and uniaxial symmetry 
and values of $g$-factor calculated for models U1-U3 according to the procedure described in the text.
In the last two columns error functions $\left<E_{RMS}^{N-1}\right>$ and $\left<E_{RMS}^{1}\right>$ 
are shown.}
\begin{tabularx}{\textwidth}{XXXXXXXXXXXX}
\hline\hline\\
Model & $H_{[001]}^{ef\!f}$ & $H_{2[001]}$ & $H_{3[001]}$ & $H_{[110]}$ & $H_{c1}$ & $H_{c2}$ & $H_{c3}$ & $H_{c4}$ & $g$ & 
$\left<E_{RMS}^{N-1}\right>$& $\left<E_{RMS}^{1}\right>$ \\
\\
\hline
\\
  U1  & 4811 &       &      & 66.28 & 78.07 & -534.1 & 43.93 & 1405 & 1.985 & 0.59 & 0.50 \\
  U2  & 4779 & 21.83 &      & 66.08 & 80.12 & -449.3 & 40.76 & 1222 & 1.987 & 0.59 & 0.51 \\
  U3  & 4778 & 21.99 & 6310 & 66.13 & 80.15 & -449.3 & 40.59 & 1221 & 1.987 & 0.59 & 0.50  \\
\\
  \hline\hline
\end{tabularx}
\label{Uni_Fields}
\end{table*}
The predictive ability, measured by the error function $\left<E_{RMS}^{1}\right>$, for all models 
defined  in Tables \ref{Cubic_present} and \ref{Uni_present}
is shown in Fig. \ref{Cr_Va_fig}.
$\left<E_{RMS}^{1}\right>$ decreases for C1 - C4 models and remains roughly constant for the C4 - C6 models.
It means that higher order terms used in free energy expansion~(\ref{field_eq})
for C5 and C6 models do not improve their ability to predict cubic magnetocrystalline
anisotropy on the basis of experimental data from Fig.~\ref{experimental_data}.
Similarly we see that taking into account higher order terms in uniaxial 
anisotropy expansion for models U2 and U3 does not improve their predictive ability.
It follows that  the correct description of magnetocrystalline anisotropy
in (GaMn)As 
requires that
the free energy expansion should include terms describing cubic anisotropy
up to fourth order and that is enough to take into account uniaxial anisotropies up to first order.
\begin{figure}[h!]
\centering
   \includegraphics[width=0.45\textwidth, angle=0]{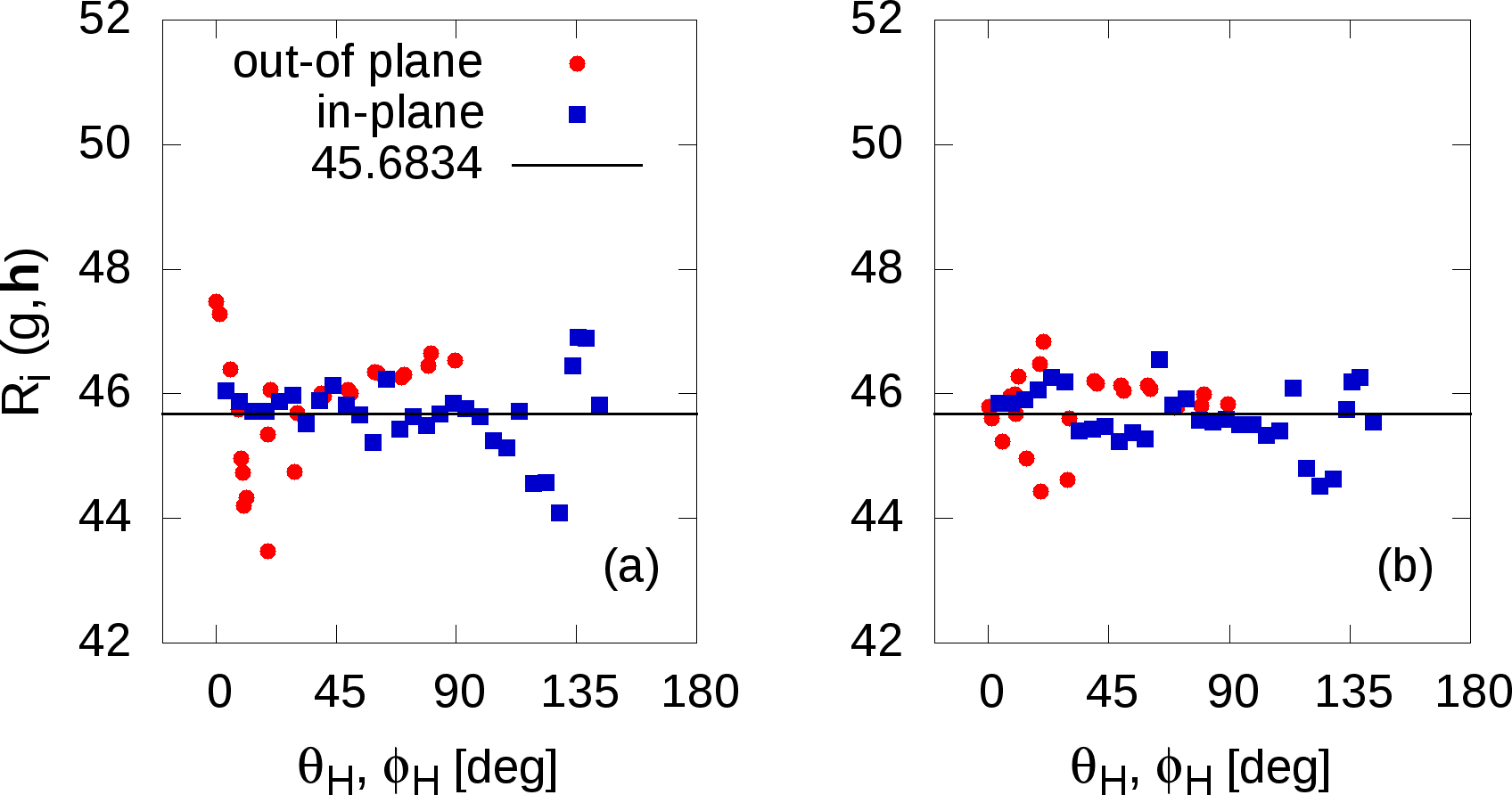}
   \caption{The values of function $R_i(g,{\mathbf h})$ defined in Eq.~(\ref{Left_SB_num_eq}) 
   for model C1 - (a) and for model C4 - (b)
   for all experimental
   points from Fig. \ref{experimental_data}. Squares -- in-plane geometry: $\vartheta_H=90^\circ$, $\varphi_H$ changes;
   circles -- out-of-plane geometry: $\vartheta_H$ changes, $\varphi_H=-45^\circ$. The values of $R_i(g,{\mathbf h})$
   were calculated for the corresponding field values taken from the Table \ref{Aniso_Fields}. }
   \label{Smit_AB}
\end{figure}

One can also see the
result of minimization in Fig. \ref{Smit_AB} in the form of a collapse: 
for the real minimum of error function
at $g^*\!, {\mathbf h}^*$
its values $R_i(g^*\!,{\mathbf h}^*)$ for all experimental points
fall onto a line 45.6834. 
By comparing the scattering of points for C1 model (a) and C4 model (b) 
we note the important thing: the addition of the higher orders of cubic anisotropy
fields improves fitting not only for {\em in-plane} experimental points but also 
for {\em out-of-plane} experimental points. 
This is due to the occurrence of partial mixed derivatives 
of the free energy in the determinant from Hessian
(representing a local curvature of free energy in $\vartheta_H, \varphi_H$ space)
in Smit-Beljers equation~(\ref{SB_eq})
which we solve numerically\footnote{We use {\em Python} packages {\em scipy.optimize}
and {\em Num\-difftools}} for all experimntal points simultaneously treating 
them on equal footing.
Therefore, 
to get as accurate as possible anisotropy field values
it is important to measure resonance fields in different geometries.

\begin{figure}[!ht]
\centering
   \includegraphics[width=0.4\textwidth, angle=0]{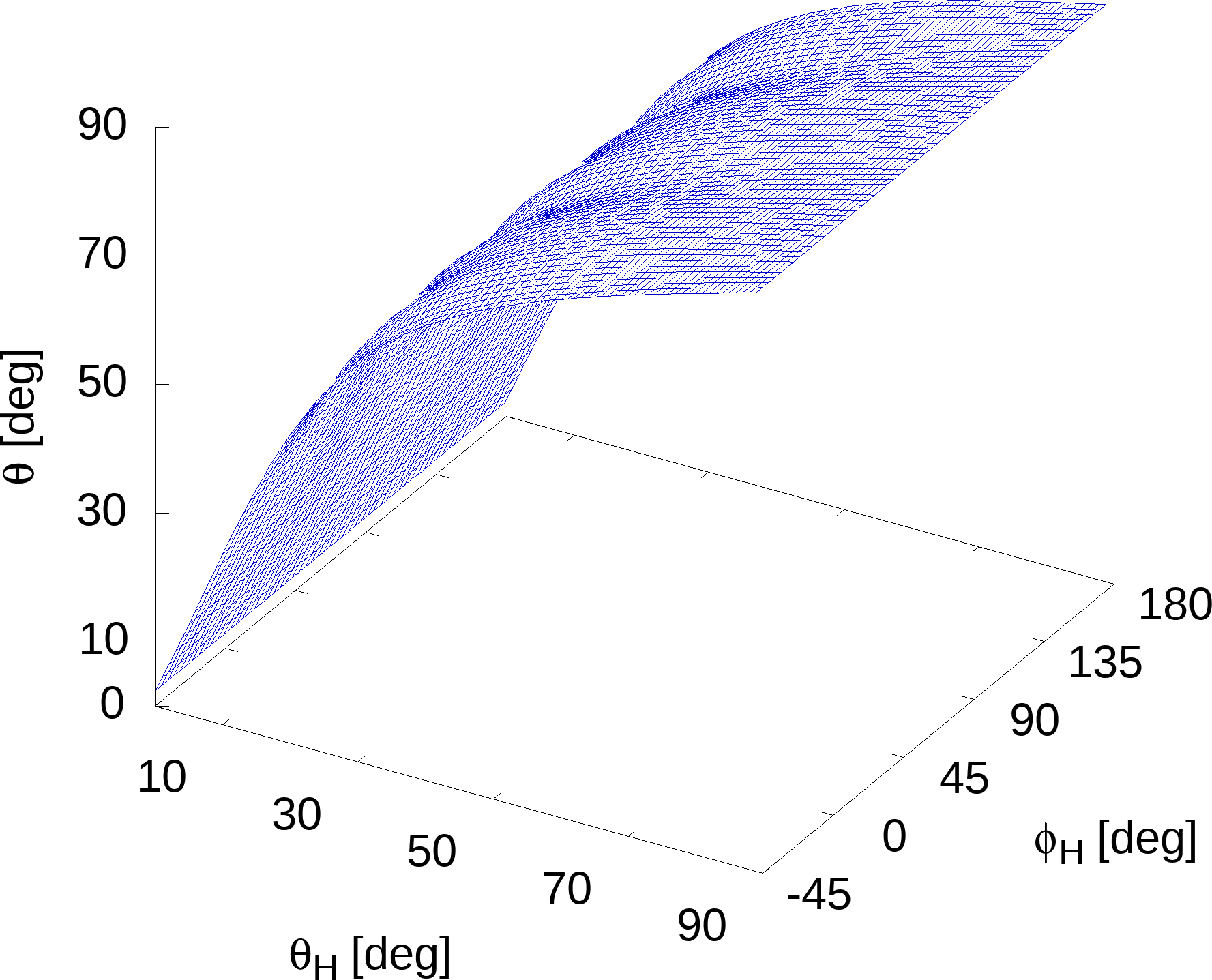}
   \caption{
    Equilibrium magnetization angle $\vartheta$ versus the external field angles
    $\vartheta_H, \varphi_H$ determined from the condition  (\ref{equilibrium_eq}) for the (Ga,Mn)As thin film
    studied in Ref. \onlinecite{Liu2007}.
   }
   \label{Angle_Theta_M}
\end{figure}

\begin{figure}[!ht]
\centering
   \includegraphics[width=0.4\textwidth, angle=0]{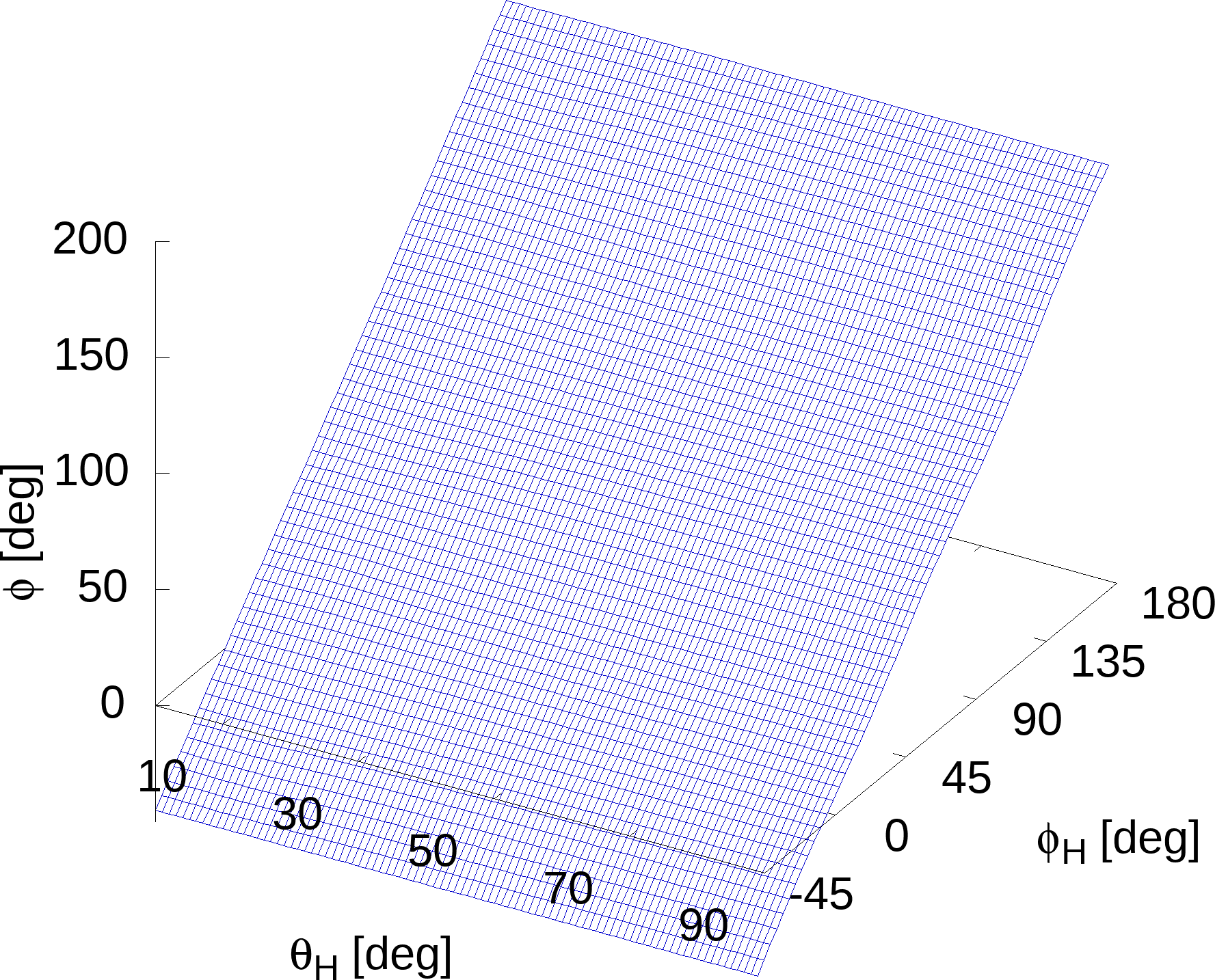}
   \caption{Equilibrium magnetization angle $\varphi$ versus the external field angles
    $\vartheta_H, \varphi_H$ determined from the condition  (\ref{equilibrium_eq}) for the (Ga,Mn)As thin film
    studied in Ref. \onlinecite{Liu2007}.}
   \label{Angle_Phi_M}
\end{figure}

Fulfilling condition (\ref{equilibrium_eq}) while solving Eq.~(\ref{Opt_num_eq}) 
leads to finding dependences $\vartheta(\vartheta_H, \varphi_H)$ and $\varphi(\vartheta_H, \varphi_H)$.
for all models from Tables \ref{Aniso_Fields} and  \ref{Uni_Fields}.
They are shown for C4 model in Figs. \ref{Angle_Theta_M} and \ref{Angle_Phi_M}.
Function $\vartheta(\vartheta_H, \varphi_H)$ for a given $\varphi_H$ it 
always is a concave function.
For angle $\phi_H=45^{\circ}$ and $135^{\circ}$ we see a ripple, which is the result of the presence of cubic symmetry.
Note also, that function $\varphi(\vartheta_H, \varphi_H)$ for a given $\vartheta_H$ is for all $\vartheta_H$
a linear function $\varphi \propto \varphi_H$ (does not depend on $\vartheta_H$).

Let us summarize this section by stating that for the correct description of magnetocrystalline anisotropy
in (GaMn)As, 
the free energy expansion should include terms describing cubic anisotropy
up to fourth order and that is enough to take into account uniaxial anisotropies up to first order.

\section{\label{Resonance Field} 
Back to the experiment: what is the effect of incorporating of anisotropy fields of higher orders?}

Let us now 
examine the dependence of the resonance field $H_r$ on $\vartheta_H$ and $\varphi_H$.
The problem can be stated in the following way: Given the values of $H_r$
on the boundary of the box presented in Fig. \ref{experimental_data}, determine 
the resonance field inside the box. One finds a solution in two stages. First, 
one determines the anisotropy fields, for which Eq.~(\ref{SB_num_eq}) is satisfied
on the boundary of the box. 
This stage has been described in the Section \ref{Resonance}.
Second,
to get the resonance field for each $\vartheta_H$ and each $\varphi_H$ one should 
solve Eq.~\ref{SB_num_eq} numerically for the anisotropy fields determined in the first stage 
(collected in Table \ref{Aniso_Fields})
with condition (\ref{equilibrium_eq}) met for each tentative point obtained during the numerical solving procedure.
\begin{figure}[!ht]
\centering
   \includegraphics[width=0.35\textwidth, angle=0]{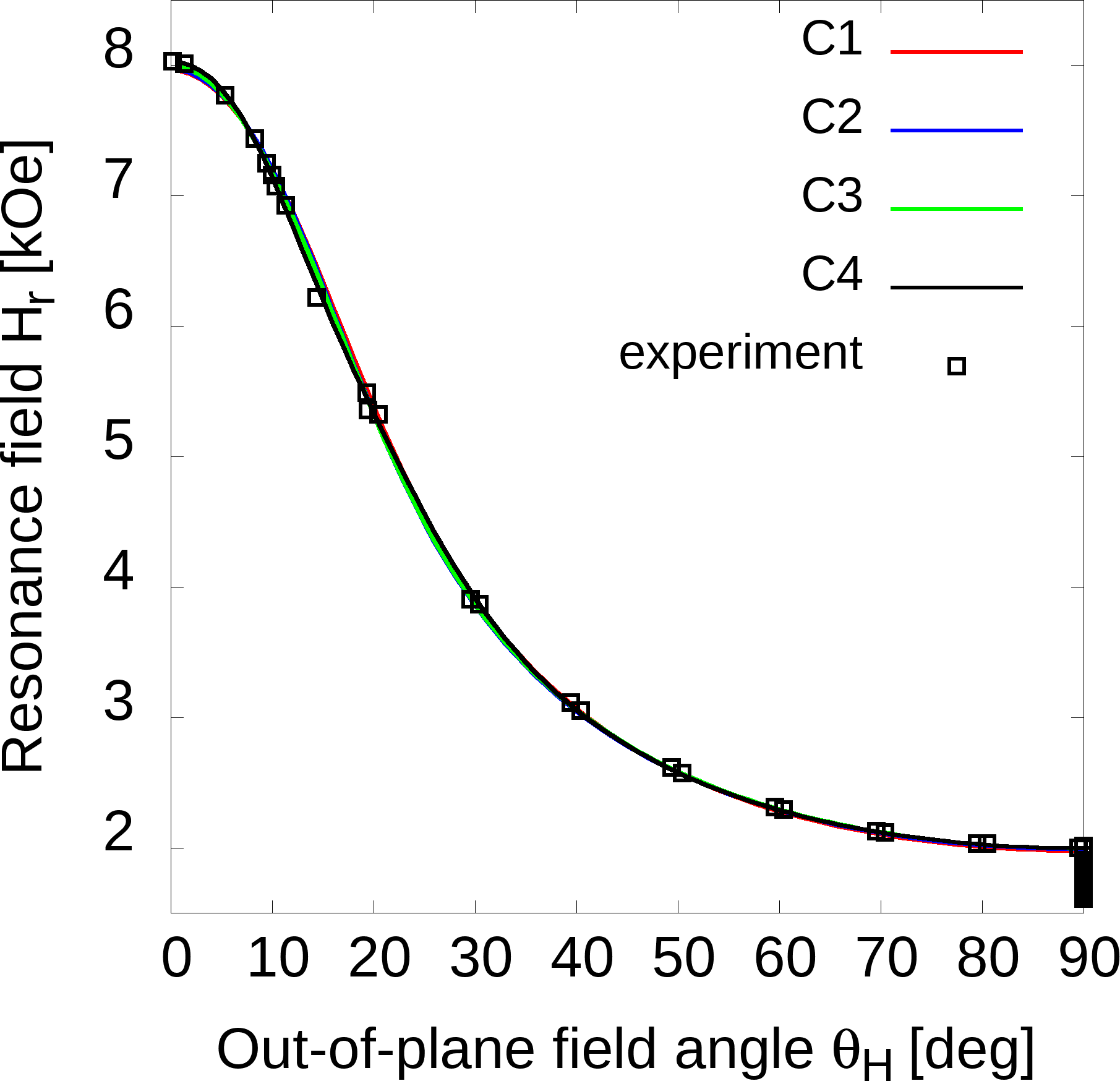}
   \caption{Comparison of the $H_r$-values calculated numerically 
            from Eq.~\ref{SB_eq} with experimental data for out-of-plane geometry for C1-C4 models.
            The resonance field for in-plane geometry measurement is visible as a~small rectangle
            on the vertical axis for $\vartheta_H=90^{\circ}$. }
   \label{Hr_Out}
\end{figure}
\begin{figure}[!ht]
\centering
   \includegraphics[width=0.35\textwidth, angle=0]{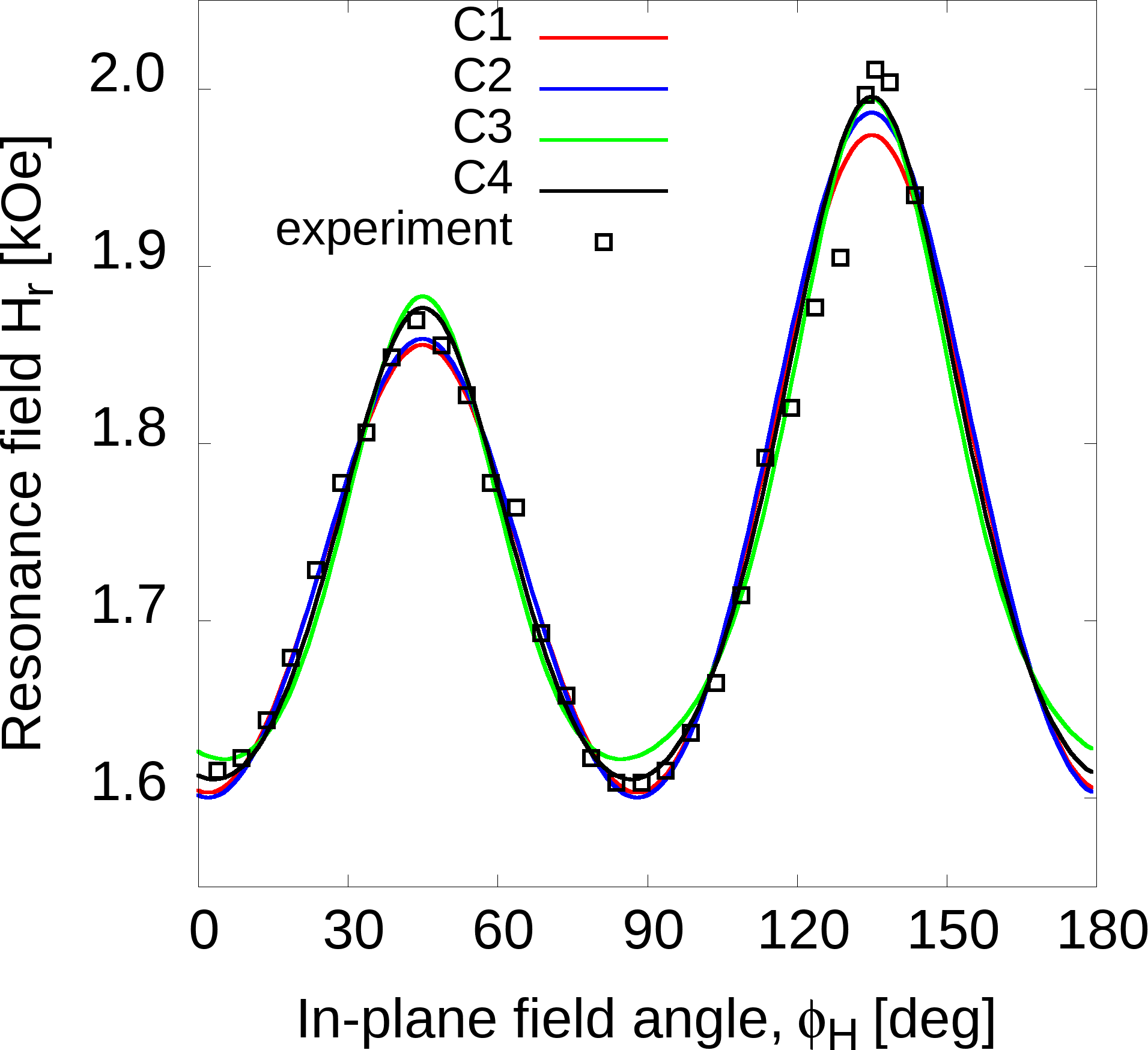}
   \caption{Comparison of the $H_r$-values calculated numerically 
            from Eq.~\ref{SB_eq} with experimental data for in-plane geometry for C1-C4 models.
            Taking into account higher order terms of cubic anisotropy one 
            improves the agreement of calculated $H_r$-values with experimental data.}   
   \label{Hr_In}
\end{figure}
In Figs.~\ref{Hr_Out} and \ref{Hr_In} one can see dependencies $H_r(\vartheta_H, -45^{\circ})$,
i.e, for out-of-plane geometry and $H_r(90^{\circ}, \varphi_H)$ -- for in-plane geometry, respectively.
Uniaxial anisotropy is the most visible for out-of-plane geometry (Fig. \ref{Hr_Out})
and although the use of higher order cubic terms does improve the agreement
of the calculated $H_r$-values with the experimental data, this improvement is not particularly visible
in the scale of Fig.~\ref{Hr_Out} because the cubic anisotropy is much smaller than the uniaxial one.
The improvement, however, can be seen for in-plane geometry:
the use of higher order terms in of cubic anisotropy expansion given by Eq.~(\ref{cubic_eq}) becomes necessary 
to  describe dependence $H_r(90^{\circ}, \varphi_H)$ more precisely.

\begin{figure}[!ht]
\centering
   \includegraphics[width=0.40\textwidth, angle=0]{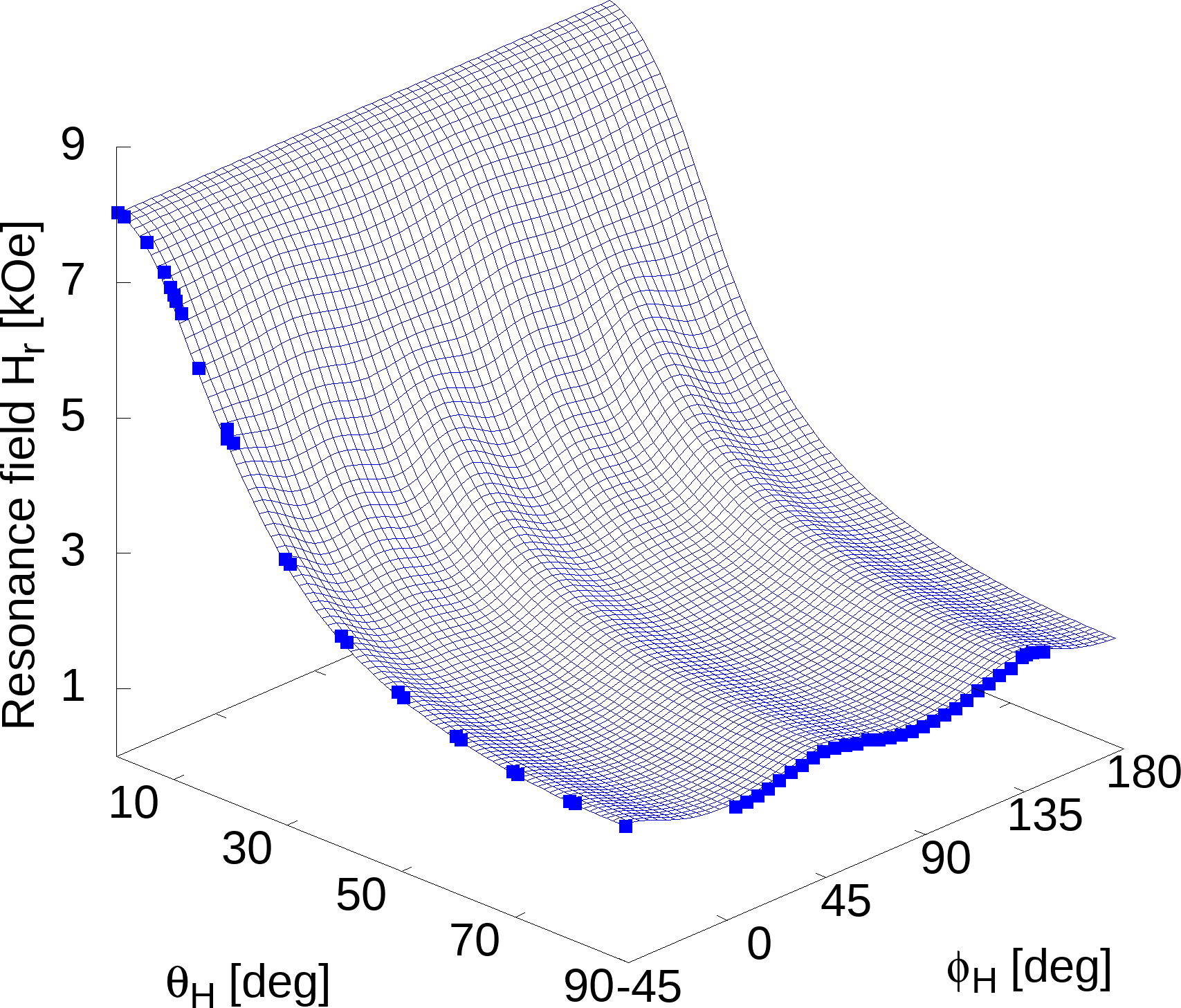}
   \caption{
   Spatial angular dependence of resonance field $H_r(\vartheta_H, \varphi_H)$
   resulting from our theory. 
   Experimental data (squares) in the  plane  $(H_r, \vartheta_H)$
   correspond to out-of-plane geometry
   while those in the $(H_r, \varphi_H)$ plane -- to in-plane geometry.
   The surface in the figure is the numerical solution of Eq.~(\ref{SB_num_eq}) 
   with condition (\ref{equilibrium_eq}).
   }
   \label{H_res}
\end{figure}

Spatial dependence of the resonance field on angles $\vartheta_H$  and $\varphi_H$
is shown in Fig. \ref{H_res}.
For small angle $\vartheta_H$ we see resonance field
whose source is 
mainly uniaxial [001] anisotropy (with two-fold symmetry), 
whereas for angle $\vartheta_H \approx 90^{\circ}$
the resonance field with four fold symmetry becomes more noticeable.
It is the result of cubic anisotropy,
although $H_r(90^{\circ}, 45^{\circ}) < H_r(90^{\circ}, 135^{\circ})$
due to small uniaxial [110] anisotropy, see also Fig. \ref{Hr_In}.

\begin{figure}
\centering
   \includegraphics[width=0.4\textwidth, angle=0]{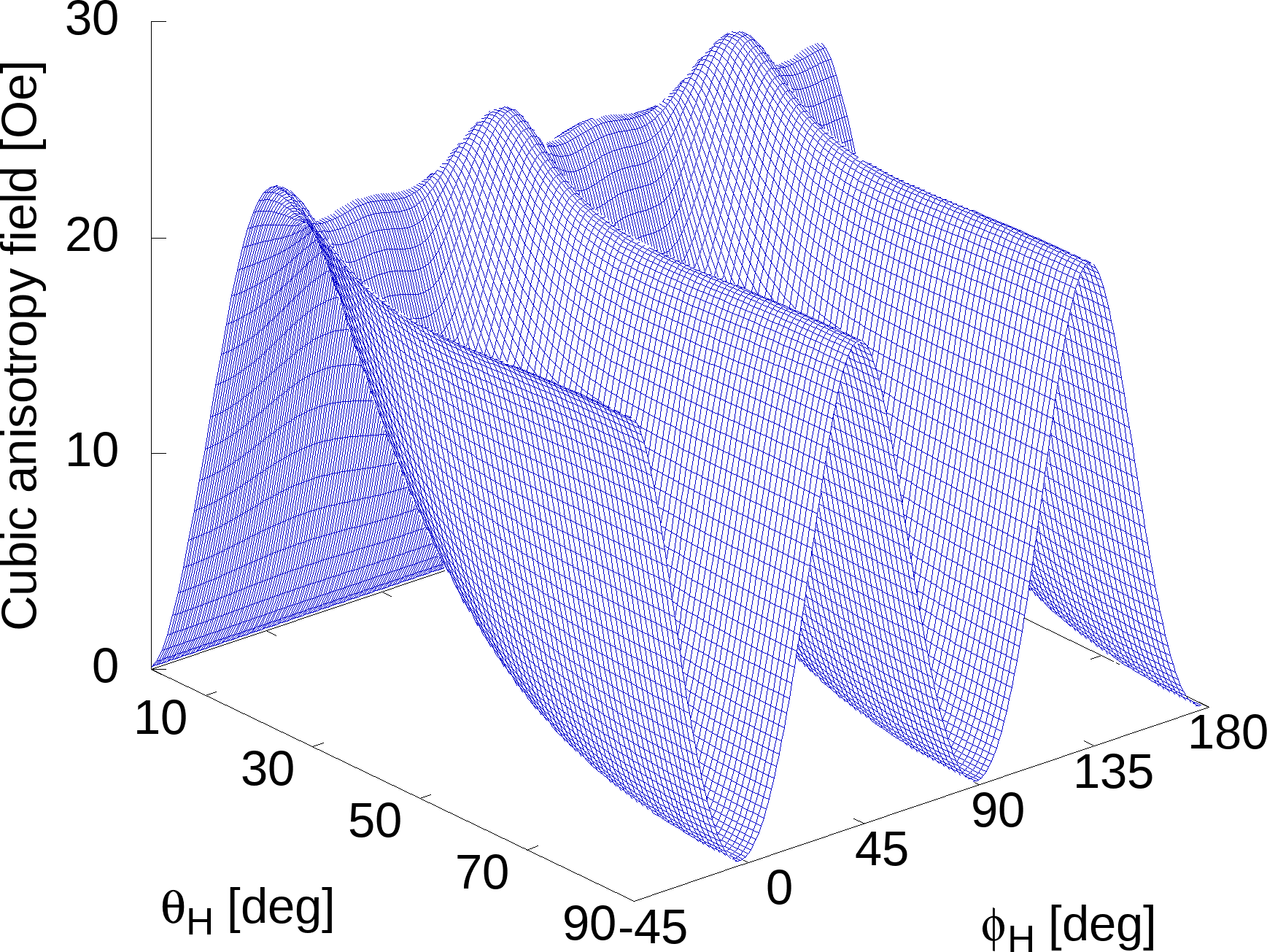}
   \caption{
           Cubic anisotropy field for C4 model versus $\vartheta_H$ and $\varphi_H$.}
   \label{Cryst_Cub}
\end{figure}

The spatial dependence of  the cubic anisotropy field is shown in the Fig.~\ref{Cryst_Cub}.
We see that really the magnetic field determines hard/easy directions, not hard/easy axes:
The largest cubic anisotropy field is for $\vartheta_H=23.7^{\circ}$ and $\varphi_H=45^{\circ}$
whereas the position of hard axis direction is given by
 $\vartheta_H=54.7^{\circ}$ and $\varphi_H=45^{\circ}$.

The values of fictitious anisotropy fields are important in that 
they allow to reproduce the spatial dependence of the energy of the sample 
in a magnetic field due to magnetocrystalline anisotropy and to find easy
and hard axes.
To find this spatial dependence one needs to know the saturation magnetization.
Then the
magnetic anisotropy constants can be easily expressed by corresponding 
anisotropy fields, see e.g., very clearly written Ref. [\onlinecite{Gutow2013}]. 
We have found the saturation magnetization\footnote{Details of 
this calculation will be published in a separate paper} of the considered sample: it amounts  $M_s=30.5\,\,\mbox{emu/cm}^3$,
 which is a typical value for (Ga,Mn)As containing a few percent of Mn atoms.
 
Returning to the Eq.~(\ref{cubic_eq}) and multiplying it by $M_s$ one obtains the spatial distribution
of energy $F_C(\vartheta, \varphi)$ stored in bulk (Ga,Mn)As and related to its cubic magnetocrystalline anisotropy
for the C4 model
 \begin{equation}
\begin{split}
F_C(\vartheta, \varphi)=
M_sH_{c1}(n^2_x n^2_y + n^2_y n^2_z + n^2_z n^2_x) + \\ 
M_s H_{c2}(n^2_x n^2_y n^2_z) + \\
M_s H_{c3}(n^4_x n^4_y + n^4_y n^4_z + n^4_z n^4_x) + \\
M_s H_{c4}(n^4_x n^4_y n^2_z + n^4_x n^2_y n^4_z + n^2_x n^4_y n^4_z) \equiv \\
K_{c1}(n^2_x n^2_y + n^2_y n^2_z + n^2_z n^2_x) + \\ 
K_{c2}(n^2_x n^2_y n^2_z) + \\
K_{c3}(n^4_x n^4_y + n^4_y n^4_z + n^4_z n^4_x) + \\
K_{c4}(n^4_x n^4_y n^2_z + n^4_x n^2_y n^4_z + n^2_x n^4_y n^4_z),
\label{cubic_Ene}
\end{split}
\end{equation}
and similarly for the C1 model.
$K_{c1}$ - $K_{c4}$ stand in Eq.~(\ref{cubic_Ene}) for cubic anisotropy constants. 
Taking the numerical values of anisotropy fields $H_{c1}$ - $H_{c4}$ from Table~\ref{Aniso_Fields}  
one obtains the numerical values of anisitropy constants for C1 and C4 models -- they are 
collected in Table~\ref{Anizo_Constants}.
Let us note that the values of first order cubic anisitropy for (Ga,Mn)As are several dozen to several hundred times smaller than 
the corresponding values for such ferromagnets as Ni or Fe.
Perhaps this is why anisotropies of higher orders become visible in resonance experiments only
for weak ferromagnets.

To assess the accuracy of the present method
we used the bootstrap method to evaluate errors for C1 and C4 models. 
To do this we assumed that the error probability distribution of experimental results was normal, and consequently 
the errors of solution of Smit-Beljers equation had also a normal distribution. Then we could determine the approximated errors 
of obtained anisotropy constants. They are listed in the Table \ref{Anizo_Constants}.

\begin{figure}
\centering
\begin{subfigure}[b]{.42\linewidth}
    \centering
    \includegraphics[width=\textwidth]{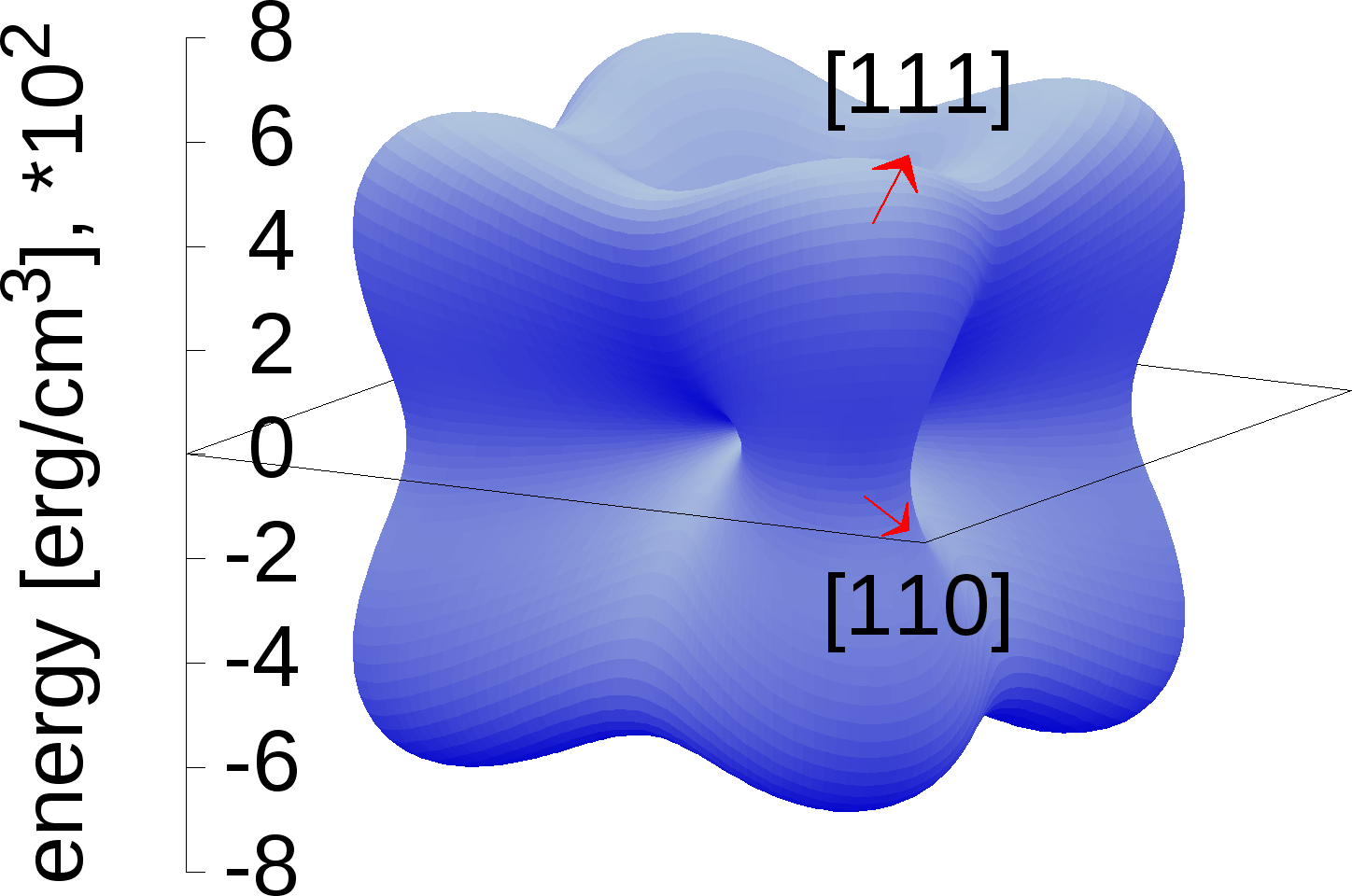}
    \caption{}\label{fig_13a}
  \end{subfigure}%
  \qquad \qquad
  \begin{subfigure}[b]{.42\linewidth}
    \centering
    \includegraphics[width=\textwidth]{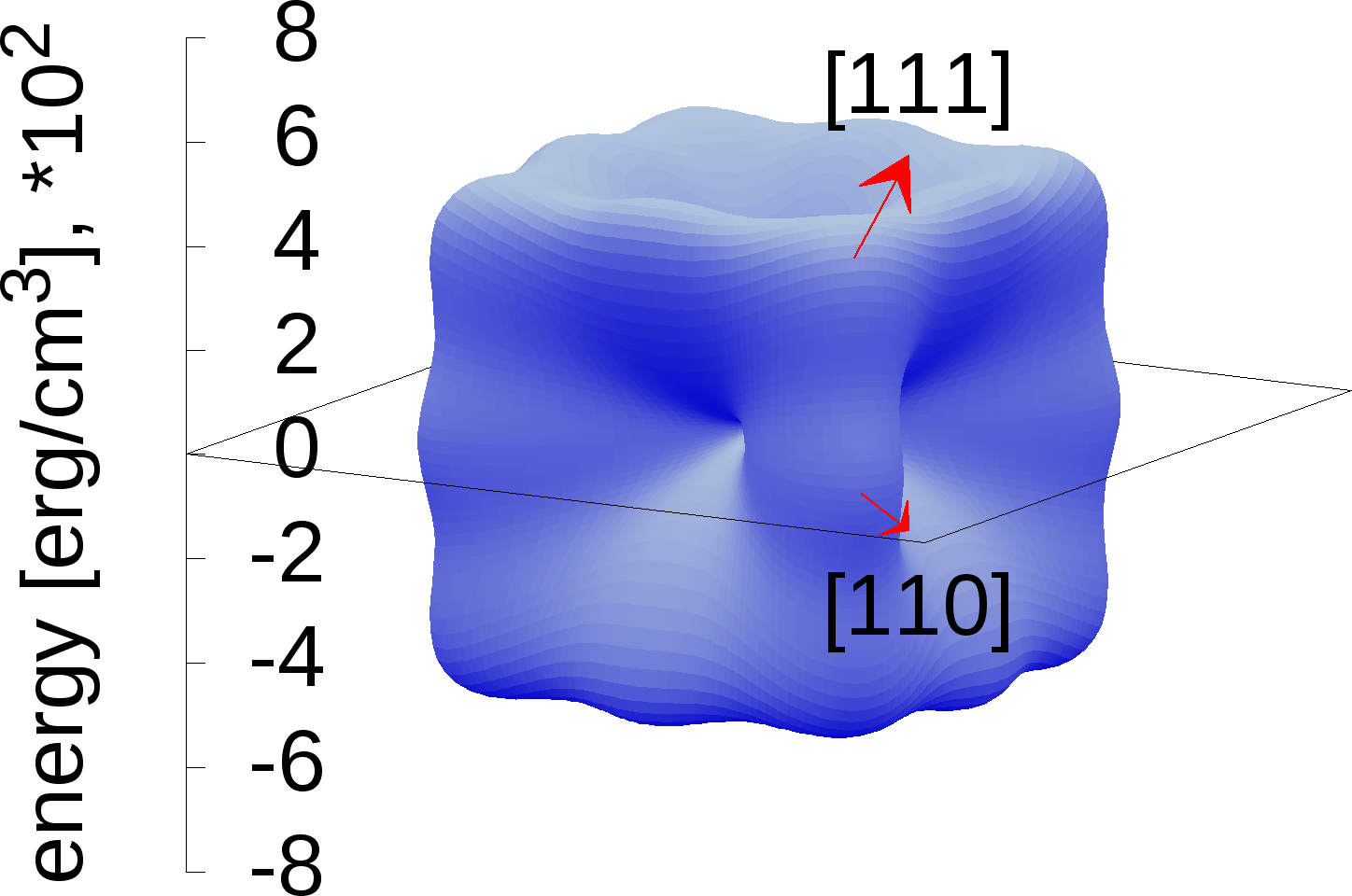}
    \caption{}\label{fig_13b}
  \end{subfigure}%
  \caption{
  Spatial dependence of cubic magnetocrystalline energy in spherical coordinate systems
  for C1 (a) and C4 (b) models.}
  \label{Cubic_Ene}
\end{figure}

\begin{figure}
\centering
   \includegraphics[width=0.3\textwidth, angle=0]{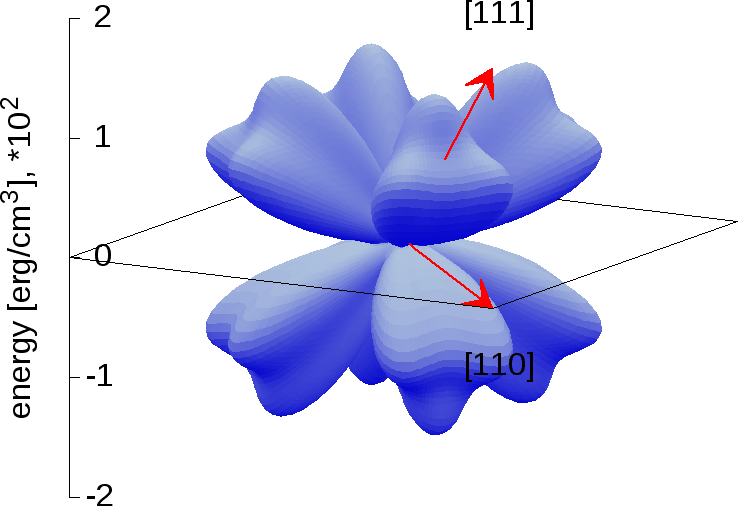}
   \caption{
  Spatial dependence of the difference of cubic magnetocrystalline energy between C1 and C4 models 
  in spherical coordinate systems. 
 }
   \label{Diff_Cubic_Ene}
\end{figure}

Finally, let us show, how taking into account the higher orders of anisotropy fields changes the cubic anisotropy energy surface.
It might seem that correction will be of little importance.
However, this is not the case: the Smit-Beljers equation~(\ref{SB_eq}) describing the curvature of energy surface
is nonlinear with respect to the second derivatives. This leads to significant corrections in energy values.
Figure~\ref{Cubic_Ene} shows the spatial dependence of energy from cubic anisotropy on the same scale for models C1 and C4.
Axes [100], [010] and [001] are easy axes with respect to cubic anisotropy and axis [111] is a hard one.
For example, for hard axis we have $F_{C1}$([111]) = 933$\pm$6, $F_{C4}$([111]) = 769$\pm$4 [erg/cm$^3$].
The C1 model thus overestimates the anisotropy energy along hard axis by about 20\%.
Figure~\ref{Diff_Cubic_Ene} shows the spatial dependence of energy difference between C1 and C4 models.

\begin{table*}
\caption{Cubic anisotropy fields ($H_{c}$) in Gaussian units [Oe] and SI units [kA/m] 
and cubic anisotropy constants ($K_{c}$) in [erg/cm$^3$] and [J/m$^3$]
for bulk (Ga,Mn)As 
calculated for models C1 and C4.
}
\begin{tabularx}{\textwidth}{XXXXXXXXX}
\hline\hline\\
     & $H_{c1}$&$H_{c2}$& $H_{c3}$& $H_{c4}$ & $K_{c1}$ &$K_{c2}$ & $K_{c3}$ & $K_{c4}$  \\
     \\
     Model C1\\
      \hline
      \\
     Gaussian& 91.81$\pm$.55 &        &       &            &2800$\pm$17&      & &  \\
     SI      & 7.306$\pm$0.044 &        &       &            &280.0$\pm$1.7&      & &  \\
     \\
     Model C4\\ 
     \hline
     \\
     Gaussian& 78.07$\pm$.40 &  -534$\pm$28   &43.9$\pm$1.2 &1410$\pm$70
             & 2381$\pm$13   &-16280$\pm$860  &1330$\pm$37  & 42900$\pm$220 \\
     SI      &6.213$\pm$0.032 &-42.5$\pm$2.3  &3.493$\pm$0.096  &112.2$\pm$5.6
             &238.1$\pm$1.3    &-1628$\pm$86   &133.0$\pm$3.7    & 4290$\pm$22 \\
\\
  \hline\hline
\end{tabularx}
\label{Anizo_Constants}
\end{table*}
\section{\label{Summary}Summary and outlooks}

The article presents how to determine bulk magnetocrysralline anisotopy 
in (Ga,Mn)As thin film by numerical solution of the Smit-Beljers equation
for all data 
collected in one FMR experiment, i.e.,
for different spatial 
orientations of the magnetic field
with respect to the sample,
on equal footing.
To avoid essential drawbacks of fitting procedures (lack of information
which fitted constants are relevant and possibility of overfiting) by finding anisotropy constants
we cross-validated the numerical solutions of Smit-Beljers equation for six models (C1-C6).
The results of this cross-validation, i.e., the values of the function $\left<E_{RMS}^{1}\right>$ displaying
predictive  ability for models C1-C6  point that it is necessary to expand
bulk cubic anisotropy up to the fourth order to reproduce spatial dependence of the resonance field correctly ---
that is increasing the order of expansion of anisptropy does not change predictive  ability of the model
under consideration.
Such cubic anisotropy (up to fourth order) is visible in the resonant experiment.
It means that
the models of first order cubic anisitropy applied so far to (Ga,Mn)As
overstimated
the value of this anisotropy.
Let us stress that this description of the bulk
anisotropy is consistent with the presented earlier 
description\cite{Pusz2017} of the surface anisotropy (both descriptions require higher order 
expansion of cubic anisotropy).
We also have shown that 
FMR data allow one
to find the spectroscopic splitting factor
with high accuracy.
We intend to
confirm the usfullness
of this new approach 
by applying it to other available resonance experiments
in the near future.

\noindent
{\bf Acknowledgment} The authors would like to thank Marcin Tomczak 
for stimulating discussions. This study is a part of a project financed by Narodowe
Centrum Nauki (National Science Centre of Poland),
grant no. DEC-2013/08/M/ST3/00967. Numerical calculations were performed at Pozna\'n Supercomputing and
Networking Center under Grant no. 284.
\bibliographystyle{apsrev4-1}
\bibliography{FMR_GaMnAs_Revised.bib}

\end{document}